\documentclass[twocolumn,aps,pra,superscriptaddress,longbibliography]{revtex4-1}
\usepackage{amsmath, amssymb, graphicx, setspace}
\usepackage[colorlinks=true,linkcolor=blue,citecolor=blue,urlcolor=blue]{hyperref}
\newcommand{\mathsym}[1]{{}}
\newcommand{\unicode}[1]{{}}
\DeclareMathOperator{\actan}{actan}

\begin{document}
%\title{Quantum Phases in a Hybrid Atom-Optomechanical system}
\title{Steady-state phase diagram of quantum gases in a lattice coupled to a membrane}
\author{Chao Gao}
\email[]{gaochao@zjnu.edu.cn}
\affiliation{Department of Physics, Zhejiang Normal University, Jinhua, 321004, China}
\author{Zhaoxin Liang}
\email[]{zhxliang@gmail.com}
\affiliation{Department of Physics, Zhejiang Normal University, Jinhua, 321004, China}

\date{\today}

%Abstract

\begin{abstract}
In a recent experiment [Vochezer {\it et al.,} Phys. Rev. Lett. \textbf{120}, 073602 (2018)], a novel kind of hybrid atom-opto-mechanical system has been realized by coupling atoms in a lattice to a membrane. While such system promises a viable contender in the competitive field of simulating non-equilibrium many-body physics, its complete steady-state phase diagram is still lacking. Here we study the phase diagram of this hybrid system based on an atomic Bose-Hubbard model coupled to a quantum harmonic oscillator. We take both the expectation value of the bosonic operator and the mechanical motion of the membrane as order parameters, and thereby identify four different quantum phases. Importantly, we find the atomic gas in the steady state of such non-equilibrium setting undergoes a superfluid-Mott-insulator transition when the atom-membrane coupling is tuned to increase. Such steady-state phase transition can be seen as the non-equilibrium counterpart of the conventional superfluid-Mott-insulator transition in the ground state of Bose-Hubbard model. 
Further, %when the quantum gas is in the superfluid phase, 
no matter which phase the quantum gas is in, 
the mechanic motion of the membrane exhibits a transition from an incoherent vibration to a coherent one when the atom-membrane coupling increases, agreeing with the experimental observations. Our present study provides a simple way to study non-equilibrium many-body physics that is complementary to ongoing experiments on the hybrid atomic and opto-mechanical systems. 
\end{abstract}

\maketitle

\section{Introduction}

%\subsubsection*{*ultracold atoms in optical lattices}

%(*Key points: Optical lattices are interfering lights; serve as periodic lattice potentials for atoms; can simulator Hubbard models; realized SF-MI transition.*)

At present there exist ongoing interests and significant efforts in realizing a novel kind of hybrid mechanical-atomic system ~\cite{Vochezer2018,Ritsch2018,Mann2018} consisting of a membrane
%in a single-sided optical cavity: 
in an optical cavity: 
Incident light is reflected by one mirror of the cavity on resonance and forms a standing wave in front of the cavity, in which the ultracold quantum gas can
be trapped.  The motivation behind these interests is twofold. 
First, the hybrid systems~\cite{Camerer2011,Vogell2013,Bennett2014,Vogell2015,Jockel2015,Moller2017} combining mechanical oscillators and ultracold atoms provide novel
opportunities for cooling, detection and quantum control of mechanical motion with applications in precision sensing, quantum-level signal transduction and for fundamental tests of quantum mechanics~\cite{Marquardt2009,Aspelmeyer2014,Sudhir2017,Harris2017}. 
Second, the combination of membrane and resonator amplifies the collective motions of the atoms, resulting in long range (phonon-like) interactions between atoms, in particular, giving phononic degrees of freedom to
optical lattice toolbox. This will thereby open new routes to mimic the lattice vibrations and quantum simulations of phonon dynamics in realistic solid materials.

Along this research line, very recently, a group from the University of Basel in Switzerland has successfully demonstrated a dynamic instability of hybrid systems as the evidence of collective atomic motion \cite{Vochezer2018}.
In more details, the authors in Ref.~\cite{Vochezer2018} direct a laser light into a cavity that contains a thin membrane placed in the center, and utilize the reflected light to construct a one-dimensional optical lattice potential for an ensemble of Rubidium atoms. They have observed an instability and large-amplitude collective atom-membrane oscillations,
as well as a phase delay in the global atomic back-action onto the light. The value and emphasis of the experiment in Ref.~\cite{Vochezer2018} is to undoubtedly prove the existence and importance of intrinsic opto-menchanical couplings, thus generates many natural questions on intrinsic couplings effects on the quantum phases of such a hybrid mechanical-atomic system. Motivated by Ref. \cite{Vochezer2018}, an immediate work \cite{Mann2018} has focused on the non-equilibrium quantum many-body behavior induced by the intrinsic opto-menchanical couplings. 
%Using a Gross-Pitaevski (GP)-like mean-field approach, and assuming a non-resonant light-induced atom-membrane interaction,
They have found a non-equilibrium phase transition between a localized symmetric state and a symmetry-broken state of atoms, arising as a result of the competition between the lattice and the membrane exerted on the atoms.   

However, the theoretical treatment of Ref.~\cite{Mann2018} is limited within the mean-field regime. For increased strength of optical lattice, previous investigations in the context of ultracold atoms~\cite{Fisher1989,Jaksch1998,Greiner2002}  have shown that a quantum phase transition from a superfluid (SF) phase to a Mott insulator (MI) phase occurs. In this case, quantum fluctuation 
plays a key role which requires treatment beyond the mean field. 
Thus a timely question arises as to what is the phase diagram of the hybrid mechanical-atomic system investigated in Ref. \cite{Mann2018}, in particular, when the optically-trapped ultracold atoms are in the Mott insulator phase, although reaching such a state remain experimentally challenging.

In this work, we are motivated to obtain the steady-state phase diagram of the hybrid mechanical-atomic system investigated in Ref. \cite{Mann2018} in the full parameter regions. 
Starting from the general effective Hamiltonian in~\cite{Vogell2013} describing both the atoms and the membrane,
 and assuming a mean-field expectation of the vibrational mode of the membrane,
  we rephrase the atom-membrane coupling as an emergent lattice potential,
   which is of the same periodicity as the original optical lattice, while differs to that by a quarter global phase. 
The two lattice potential compose an effective one, based on which, we obtain an effective single-band Bose-Hubbard model (BHM) for the atoms. 
Our main results can be outlined as follows:

1. Within a self-consistent mean-field theory, we identify different steady-state quantum phases. 
We find that the membrane can be either in coherent vibration (CV) or in incoherent vibration (ICV) phase, while the atoms can be either in the SF phase or in the MI phase. 
We obtain phase diagrams as some varying parameters including the lattice depth, the atomic interaction and the atom-membrane coupling. 
Due to the interplay between the atoms and the membrane, different phases of the two objects are overlapped. 
Such scenario shares some similarity in the hybrid system of  a cavity mode and intra-cavity lattice bosons, where both superradiance phase transition and SF-MI phase transition are in presence \cite{Bakhtiari2015,Klinder2015,Landig2016,Chen2016}.

%\subsubsection*{*Mann2018}
%(*key points: quantum many-body behavior; considered problem: steady state; non-equilibrium? model, effective Hamiltonian; assume BEC*)
2. The recent theoretical work~\cite{Mann2018} has also investigated such hybrid systems, while is limited within the mean-field regime.
We note that they assume the bosonic atoms are in the SF phase and use a Gross-Pitaevski (GP)-like mean-field approach. 
In the same regime, although interpreted in a different way, our results are consistent with their finding.
While we have further investigated the MI phase for the atoms, 
which is of great importance due to its strongly correlated nature. 
Meanwhile, by tuning the various controlling parameters in this system, 
we present a comprehensive analysis for the SF-MI transition. 

3. Our work together with Ref. \cite{Mann2018}  give a complete description of the steady phase diagram in both the superfluid and insulating phases. More elaborate theoretical treatments for the intermediate case are beyond the scope of this work.
%Meanwhile, the various controlling parameters in this system will certainly enrich the physical behaviors beyond those of the two separate composing systems, and thus call for a comprehensive investigation. 

%(*sth more?*)

%\subsection*{*prospect}

%(*currently I think it{'}s not necessary to be stated here; while we shall say sth in the discussion or summary part. *)

%\subsection*{*structure of this article}

The emphasis and value of the present work is to provide a theoretical model, i.e. an extended Bose-Hubbard model coupled to a quantum harmonic oscillator in describing the hybrid mechanical-atomic system with capturing the key informations of both quantum many-body physics and non-equilibrium nature.  We remark that in the case of vanishing the intrinsic opto-menchanical couplings of $\lambda$, our model can be simplified into the equilibrium Bose-Hubbard model which has been widely explored both theoretically and experimentally in the context of the ultracold quantum gas. Further note that the intrinsic opto-menchanical couplings of $\lambda$  in the experiments can be engineered, as in recent studies of Ref. \cite{Vochezer2018,Mann2018}. We hope the model adopted in this work can serve as an  alternative model to
study the non-equilibrium many-body physics as well as crossovers from equilibrium to non-equilibrium physics, in a highly controllable way.

This paper is organized as follows: we briefly describe the system and our model in Sec.~\ref{sec:model}. 
Then in Sec.~\ref{sec:MF} we deal with this system within a mean-field approach separately for the membrane and the atoms in details. 
In Sec.~\ref{sec:PD} we present the phase diagram and analyze the phase transitions. 
Finally, Sec. \ref{sec:conc} is devoted to some remarks and conclusions. 

\section{\label{sec:model}Model Hamiltonian}

%\subsection*{*system}

Our goal is to study the steady-state phase diagram of the hybrid mechanical-atomic system experimentally realized in Ref. \cite{Vochezer2018}. In more details, the model system studied in this work consists of a nano-membrane in an optical cavity,  which is out-coupled with a distant ensemble of bosonic atoms via a laser light. There, incident light is reflected on resonance, forming in front of the cavity an  optical lattice that traps the ultracold quantum gas.

We consider the atom part of our model system trapped in an optical lattice of $V^{(0)}(z)$ along the $z$- direction, whereas the model system is uniform in the $x$- and $y$- directions. In such, for 
the atom along the transverse directions is uniform, the freedom along the $x$- and $y$- directions decouples from the $z$-direction, leading to the realization of a quasi-one-dimensional geometry.
The atom part of our model system can be well described with the following second-quantized Hamiltonian
\begin{equation}
H_\text{a}=\int dz\left[ \Psi ^{\dagger }(z)H_0\Psi (z)+\frac{g}{2}\Psi ^{\dagger }\Psi ^{\dagger }\Psi \Psi \right].\label{Atom}
\end{equation}
Here $\Psi$ and $\Psi^\dagger$ denote the creation and annihilation operators for the bosonic atoms respectively;  $\int dz |\Psi|^2=N_{\text a}$ describes the normalization with $N_{\text a}$ being the total number of atoms,  and $g$ labels the s-wave interatomic interaction coupling constant. In addition, $H_0$ in Eq. (\ref{Atom}) is a single-particle Hamiltonian which
reads as  ($\hbar=1$)
\begin{equation}
%H_0=-\frac{1}{2m}\frac{\partial ^2}{\partial z^2}+V \sin ^2(k z)-\mu.\label{SingleH}
H_0=-\frac{1}{2m}\frac{\partial ^2}{\partial z^2}+V^{(0)}(z)-\mu.\label{SingleH}
\end{equation}
%The first and second terms in Eq. (\ref{SingleH}) describe the kinetic and potential energies due to optical lattice with strength $V$, respectively;
The first term in Eq. (\ref{SingleH}) describes the kinetic energy, and the second term, reading as $V^{(0)}=V \sin^2(k z)$, describes the potential energy due to an optical lattice with lattice strength $V$.
Here $k$ labels the wave vector of the laser generating the lattice, while $m$ and $\mu$ the atom mass and the chemical potential respectively. 
We denote the recoil energy of the optical lattice as $E_\text{r}=k^2/2m$. Note that the Hamiltonian (\ref{Atom}) can be derived into the well-known Bose-Hubbard model within the single-band approximation, which has been 
widely studied in the context of the ultracold quantum gas.

The motion of the membrane can be treated as a one-dimensional quantum oscillator with frequency $\Omega$, i.e.,
\begin{equation}
H_\text{m}=\Omega a^{\dagger }a,
\end{equation}
with \(a\) (\(a^{\dagger }\)) being the bosonic annihilation (creation) operator of the oscillator.
  
The coupling between the membrane and the atoms can be modeled within a Born-Markov approximation by adiabatically eliminating the light field \cite{Vogell2013,Mann2018}, which gives
\begin{equation}
\label{eq:Ham}
H_{\text{a-m}}=-\lambda \left(a^{\dagger }+a\right)\int dz \Psi ^{\dagger }(z)\sin (2k z)\Psi (z),
\end{equation}
where \(\lambda\) is the atom-membrane coupling constant. We remark that the Born-Markov approximation adopted in Eq., (\ref{eq:Ham}) can be adjusted in the bad-cavity limit, as pointed out by 
Refs. \cite{Vochezer2018,Mann2018}

Therefore, the total Hamiltonian of the coupled system can be written as $H_{\text{tot}}=H_\text{a}+H_\text{m}+H_{\text{a-m}}$. 
Before investigating the steady-state phase diagram based on the Hamiltonian $H_{\text {tot}}$, we first briefly review some important features
of the Hamiltonian $H_{\text{tot}}$. 
(i) It's clear that the total Hamiltonian preserves the $U(1)$ symmetry of the bosonic atoms, while   
(ii) the $U(1)$ symmetry of the quantum oscillator that $H_\text{m}$ preserves is explicitly broken due to coupling term, $H_\text{a-m}$; 
(ii)nevertheless, the total Hamiltonian, $H_{\text{tot}}$, preserves a combination of \(\mathbb{Z}_2\) symmetry of the membrane and a spatial inversion along the lattice direction, which is also of \(\mathbb{Z}_2\) nature.

\section{Mean-field treatment \label{sec:MF}}

The goal of this section is to derive the complete steady-state phase diagram of the model system. At the heart of our solution of non-equilibrium dynamics for
the  hybrid mechanical-atomic system is an elimination of the degrees of freedom of the membrane, leading to 
an effective Bose-Hubbard model where the parameters are significantly renormalized by the atom-membrane coupling.

As a first step, the equation of motion of the operator $a$ describing the dynamics of the membrane can be obtained by applying the Heisenberg{'}s equation, 
\begin{equation}
\label{eq:a-EoM}
i\frac{da}{dt}=(\Omega -i\gamma) a-\lambda  \int dz \Psi ^{\dagger }(z)\sin (2k z)\Psi (z),
\end{equation}
Here the \(\gamma\) represents a phenomenological damping rate. As emphasized by Ref. \cite{Mann2018},  the loss rate $\gamma$ arises from the damping of the membrane and 
the radiation pressure, for which first-principle derivation is beyond the scope of this work. 

 We proceed to solve Eq. (\ref{eq:a-EoM}) following the standard procedures~\cite{Mann2018} within the mean-field framework. We approximate the field operator $a$ as
\begin{equation}
\label{eq:aEoM}
a\simeq \langle a\rangle \equiv \sqrt{N_\text{L}}\alpha ,
\end{equation}
with \(N_\text{L}\) being the number of lattice sites and \(\alpha \equiv \alpha _1+i \alpha _2\) a complex number ($\alpha _1$ and $\alpha _2$ being its real and imaginary part respectively). We point out that the $\alpha$ in Eq. (\ref{eq:aEoM}) acts as the order parameter for the membrane. In more details, \(\alpha =0\) denotes an incoherent vibration (ICV) state of the membrane, whereas \(\alpha \neq 0\) denotes a coherent vibration (CV). Plugging Eq.~(\ref{eq:aEoM}) into Eq. ~(\ref{eq:a-EoM}), we can obtain
\begin{equation}
i\frac{d\alpha }{dt}=(\Omega -i\gamma) \alpha -\Lambda  S,
\end{equation}
where $\Lambda =\lambda \sqrt{N_\text{L}}$ is a renormalized coupling constant, and
\begin{equation}
\label{eq:S}
S=\frac{1}{N_\text{L}}\int dz \Psi ^{\dagger }(z)\sin (2k z)\Psi (z)
\end{equation}
is a Hermitian quantity. Note that the $S$ essentially quantifies the interplay between the atom and the membrane, whose physical meaning will be clarified later. 

We are interested in the case when the membrane is in a steady state: \(d\alpha /dt=0\) and obtain 
\begin{equation}
\label{eq:aS}
\alpha =\frac{\Lambda  S}{\Omega -i\gamma},
\end{equation}
from which it can be inferred that \(\alpha\) is in concurrence with \(S\), and actually they are proportional to each other. Further, we can decompose
\(\alpha\) in terms of its real and imaginary parts as $\alpha _1=\Lambda  S/\tilde{\Omega}$ and $\alpha _2=(\gamma/\Omega) \alpha_1$ with $\tilde{\Omega}=(\Omega^2+\gamma^2)/\Omega$.
 One can notice that $\alpha _1$ and $\alpha _2$ are dependent on each other.

Next, by plugging Eq. (\ref{eq:aS}) into Hamiltonian~(\ref{eq:Ham}) , we can rewrite the atom-membrane coupling term as follows
\begin{equation}\label{eq:MFAM}
H_{\text{\text{a-m}}}^{\text{MF}}=-2\Lambda  \alpha _1\int dz \Psi ^{\dagger }(z)\sin (2k z)\Psi (z).
\end{equation}
Based on Eq. (\ref{eq:MFAM}), we conclude that the back action of the membrane on the optically-trapped quantum gas provides an effective optical potential in form of
\begin{equation}
V^{(1)}=-2\Lambda  \alpha _1\sin (2k z).
\end{equation}
Here, we would like to rewrite the total periodic potential subject to the quantum gas as  as follows
\begin{eqnarray}\label{eq:Veff}
V^{\text{eff}}&=&V^{(0)}+V^{(1)}\nonumber \\
&=&\sqrt{V^2+\left(4\Lambda  \alpha _1\right)^2}\sin ^2[k(z-\delta)]\nonumber\\
&-&\frac{\sqrt{V^2+\left(4\Lambda  \alpha _1\right)^2}-V}{2},
\end{eqnarray}
with
$\delta=(1/2k)\actan\left(4\Lambda  \alpha _1/V\right)$. Two properties of the
effects of the back action of the membrane on the quantum gas can immediately be stated based on Eq. (\ref{eq:Veff}): (i)
this effective lattice with the renormalized lattice strength shares the same periodicity as the original optical lattice \(V^{(0)}\) with lattice unit length \(a_\text{L}=\pi /k\);
(ii) its lattice site location is shifted from that of the original lattice, \(z_i^{(0)}=i a_\text{L}\) (\(i=0,1,2\text{...}\)), to \(z_i=i a_\text{L}+\delta\) by \(\delta\). 
The back action of the membrane on the quantum gas is to provide the competition of the lattice, trying to localize the atoms at the minima, and the membrane displacement which tries to shift the atoms.
The relation between $\delta$ and $\alpha_1$ imply that the onset of the lattice shift and the CV order of the membrane occurs simultaneously and their signs are in accordance. 
Meanwhile, as long as the two quantities are nonzero,
 the aforementioned $\mathbb{Z}_2\times \mathbb{Z}_2$ symmetry of the membrane and the spatial inversion breaks.

Finally, the total Hamiltonian of our model system can be effectively rewritten as
\begin{equation}
\begin{aligned}
H^{\text{eff}}_{\text{tot}}=&\int dz \Psi ^{\dagger }\left(-\frac{1}{2m}\frac{d^2}{dz^2}+V^{\text{eff}}-\mu \right)\Psi \\
&+\frac{g}{2}\int dz \Psi ^{\dagger }\Psi ^{\dagger }\Psi \Psi
+\Omega N_\text{L}\left|\alpha\right|^2.
\end{aligned}
\end{equation}
With the single-band approximation, one can then proceed to expand the field operator \(\Psi (z)=\sum _i w_i(z)b_i\) in terms of the annihilation operator \(b_i\) at lattice site \(i\) and the
Wannier function \(w_i(z)\equiv w\left(z-z_i\right)\), corresponding to the effective lattice \(V^{\text{eff}}\). 

Following the standard procedures as in Refs. \cite{Jaksch1998,Bloch2008}, we can then obtain an effective single-band Bose-Hubbard model (BHM):
\begin{equation}
\label{eq:BHMeff}
\begin{aligned}
H_{\text{BH}}^{\text{eff}}=&
\Omega N_\text{L}\left|\alpha\right|^2
-t \sum _i \left(b_i^{\dagger }b_{i+1}+b_{i+1}^{\dagger }b_i\right)\\
&+\frac{U}{2}\sum _i n_i\left(n_i-1\right)-\tilde{\mu}\sum _i n_i,
\end{aligned}
\end{equation}
where
\(t=-\int dz w_i^*\left[-(1/2m)d^2/dz^2+V^{\text{eff}}\right]w_{i+1} \) is the hopping parameter, 
%\begin{equation*}
%U=g\int dz \left| w_i(z)\right| ^4
%\end{equation*}
\(U=g\int dz \left| w_i\right| ^4\) is the on-site interaction energy, and
%\begin{equation*}
%\tilde{\mu }=\mu -\epsilon ^{(0)}
%\end{equation*}
\(\tilde{\mu }=\mu -\epsilon ^{(0)}\) is the effective chemical potential with 
%\begin{equation*}
%\epsilon ^{(0)}=\int dz w_i^*(z)\left[-\frac{1}{2m}\frac{d^2}{dz^2}+V^{\text{eff}}(z)\right]w_i(z)
%\end{equation*}
\(\epsilon ^{(0)}=\int dz w_i^*\left[-(1/2m)d^2/dz^2+V^{\text{eff}}\right]w_i\) being the energy of the first Bloch band.

The physics of the extended Bose-Hubbard model in Eq. (\ref{eq:BHMeff}) is determined by three basic parameters: the tunneling rate $t$,  interaction strength $U$ and the chemical potential$\tilde{\mu}$. In the tight-binding limit of \(\tilde{V}\gg E_\text{r}\), one can properly approximate \cite{Zwerger2003} the Wannier function as the Gaussian ground state in the local potential well around each site: 
$w_i(z)\approx 1/(\pi ^{1/4}d^{1/2})\exp \left[-(z-z_i)^2/4d^2\right]$,
with \(d=\left(4m^2 \tilde{V}E_\text{r}\right){}^{-1/4}.\) 
Hence, the three parameters of the extended Bose-Hubbard model in Eq. (\ref{eq:BHMeff}) can be immediately calculated analytically
\begin{eqnarray}
t&=&\frac{4E_\text{r}}{\sqrt{\pi }}\left(\frac{\sqrt{V^2+\left(4\Lambda  \alpha _1\right)^2}}{E_\text{r}}\right)^{\frac{3}{4}}e^{ -2\left(\frac{\sqrt{V^2+\left(4\Lambda  \alpha _1\right)^2}}{E_\text{r}}\right)^{\frac{1}{2}}},\label{Parametert}\\
U&=&\frac{g k}{\sqrt{2\pi}}\left(\frac{\tilde{V}}{E_\text{r}}\right)^{\frac{1}{4}},\label{ParameterU}\\
\epsilon ^{(0)}&=&\left(\sqrt{V^2+\left(4\Lambda  \alpha _1\right)^2}E_\text{r}\right)^{\frac{1}{2}}-\frac{\sqrt{V^2+\left(4\Lambda  \alpha _1\right)^2}-V}{2}.\label{Parametere}
\end{eqnarray}

At the mean-field level \cite{Fisher1989,Sheshadri1993}, we approximate the field operator \(b_i\) as a complex number
\begin{equation}\label{beta}
b_i\simeq \langle b_i\rangle \equiv \beta ,
\end{equation}
which serves as the order parameter for the bosonic atoms: \(\beta \neq 0\) ( \(\beta =0\)) denotes a superfluid (SF) [Mott-insulator (MI)] state
of the atoms. Then the effective BHM, (\ref{eq:BHMeff}), becomes $H^\text{MF}_\text{tot}=\sum_i{h_{\text{eff}}^{\text{MF}}}$, with
\begin{equation}
\label{eq:BHMmf}
h_{\text{eff}}^{\text{MF}}=-2 t \left(\beta ^* b+\beta  b^{\dagger }\right)+\frac{U}{2}n(n-1)-\tilde{\mu } n
+\Omega \left|\alpha\right|^2,
\end{equation}
where \(n=b^{\dagger }b\) is the atom number operator at each site. 
Within the mean-field approach, we can self-consistently solve for the SF order parameter $\beta $ with the filling number \(\rho =\langle n\rangle\).
On the other hand, the quantity $S$ in Eq.~(\ref{eq:S}) can be also expressed in terms of the Wannier function as follows
\begin{equation}\label{eq:Smf}
S=\frac{1}{N_\text{L}}\sum _{i,j} s_{i,j}b_i^{\dagger }b_j,
\end{equation}
with $s_{i,j}=\int dz w_i^*\sin (2k z)w_j$. 

Now, we are ready to determine the steady phase of our model system by self-consistently solving Eqs. (\ref{eq:aS}), (\ref{eq:BHMmf}) and (\ref{eq:Smf}):
(i) with the initial values of the $t$, $U$, and $\tilde{\mu}$ in Eqs. (\ref{Parametert},~\ref{ParameterU},~\ref{Parametere}) and the filling number $\rho$, one can solve the Hamiltonian of Eq. (\ref{eq:BHMmf})
following the standard method in Refs. \cite{Fisher1989,Sheshadri1993}; (ii) after obtaining the value of $\beta$ and the wave functions corresponding to the Hamiltonian of Eq. (\ref{eq:BHMmf}),
one can proceed to calculate the value of $S$ by Eq. (\ref{eq:Smf}); (iii) the values of $\alpha_1$ and $\alpha_2$ can then be obtained by Eq. (\ref{eq:aS}), which will renormalize the $t$, $U$ and $\tilde{\mu}$ of the Bose-Hubbard model  through  Eqs. (\ref{Parametert},~\ref{ParameterU},~\ref{Parametere}). It's clear that the present system of the Hamiltonian (\ref{eq:BHMeff}) has the
crucial novelty of being an intrinsically non-equilibrium system compared with the previous Bose-Hubbard model in Refs.\cite{Fisher1989,Jaksch1998,Greiner2002}.

We further point out that \(S\) actually serves as an order parameter that quantifies the displacement of the atoms due to the onset of \(\alpha\)
and thus the additional emergent lattice \(V^{(1)}\). If \(V^{(1)}=0\), the lattice site is located at the minimum of \(V^{(0)}\), which is \(z_i^{(0)}=i
a_\text{L}\) (\(i=0,1,2...\)). Correspondingly the Wannier function \(w_i(z)\) is symmetric around \(z_i^{(0)}\), while \(\sin (2 k z)\) is anti-symmetric
around \(z_i^{(0)}\), then \(s_{i,j}\) always vanishes. However, if \(V^{(1)}\neq 0\), the lattice site is located at the minimum of \(V^{\text{eff}}\),
that is \(z_i=i a_\text{L}+\delta\), which is shift from \(z_i^{(0)}\) by \(\delta\). Although the Wannier function \(w_i(z)\) is
still symmetric around \(z_i\), the function \(\sin (2 k z)\) is no more anti-symmetric around \(z_i\), then \(s_{i,j}\) and correspondingly \(S\) do not vanish.

Within mean-field approximation, and classifying terms by on-site \(i=j\) or off-site \(i\neq j\), \(S\) becomes
\begin{equation*}
S=\frac{1}{N_\text{L}}\sum _{i,j} s_{i,j}\langle b_i^{\dagger }b_j\rangle 
=\frac{1}{N_\text{L}}\left(\rho\sum _i s_{i,i}+\left| \beta \right| ^2 \sum _{i\neq j} s_{i,j}\right).
\end{equation*}
Here we have used \(\langle b_i^{\dagger }b_i\rangle =\rho\),
% where \(N_\text{L}\) is the number of lattice sites, 
and \(\langle b_i^{\dagger }b_j\rangle \simeq \langle b_i^{\dagger }\rangle \langle b_i\rangle =\left| \beta \right| ^2\)
 for \(i\neq j\). 
Note that due to the hermiticity of the original definition, the mean-field value of \(S\) must be real. Due to the symmetry of lattice translation,
\(s_{i,j }\) only depends on the relative difference: \(i-j\).

Further by taking the aforementioned approximation for the Wannier function, one finds
\begin{equation*}
s_{i,j}\simeq  (-1)^{i-j}\exp \left[-\frac{(i-j)^2}{4}\left(\frac{\tilde{V}}{E_\text{r}}\right){}^{1/2}\right]s_{0,0} ,
\end{equation*}
and
\begin{equation*}
s_{0,0}\simeq  \sin (2k \delta)e^{-k^2d^2}=\frac{4\Lambda  \alpha _1}{\tilde{V}}\exp \left[-\left(\frac{E_\text{r}}{\tilde{V}}\right){}^{1/2}\right].
\end{equation*}
Note that all \(s_{i,j}\) terms are real since we have taken the Wannier functions to be real. Due to the exponentially decaying nature when \(\tilde{V}\gg
E_\text{r}\), those off-site terms \(s_{i,j}\) (\(i\neq j\)) are negligible, hence we can further approximate 
\(S\simeq \rho s_{0,0}\).
%\begin{equation}
%\label{eq:Ss}
%S\simeq s_{0,0}.
%\end{equation}
\section{Phase diagram\label{sec:PD}}

\begin{figure}[!hbp]
\includegraphics[width=8cm]{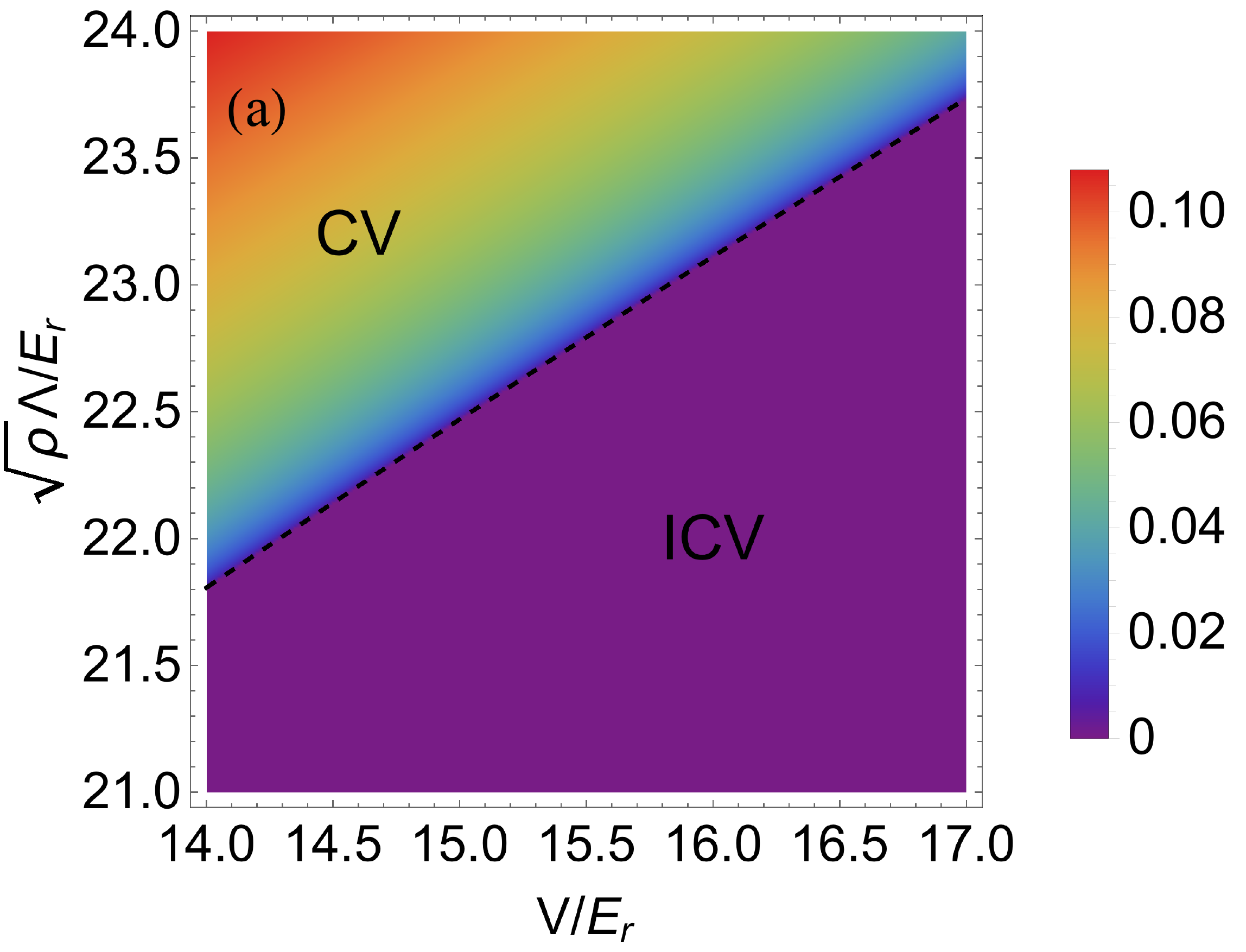}
\includegraphics[width=8cm]{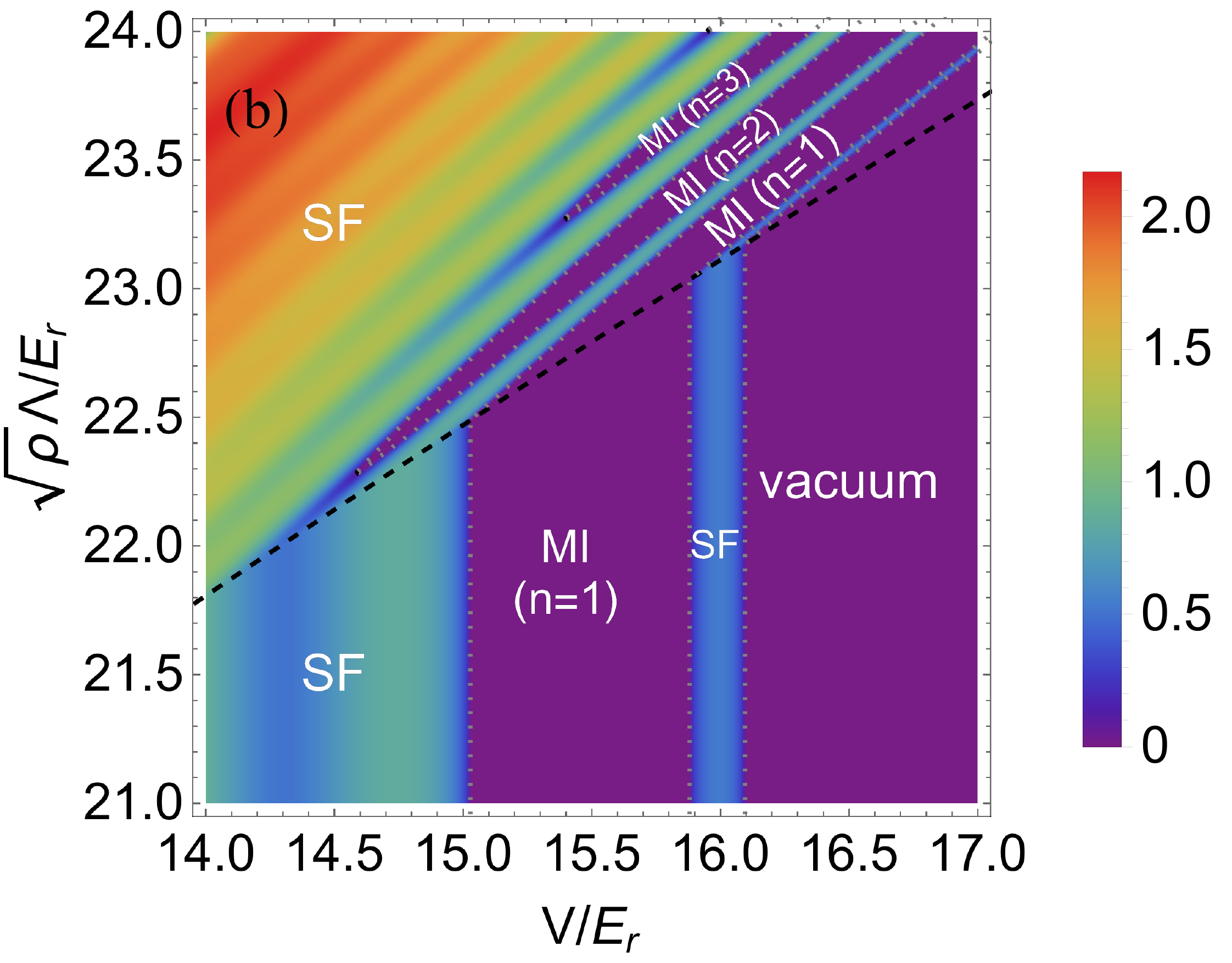}
\includegraphics[width=8cm]{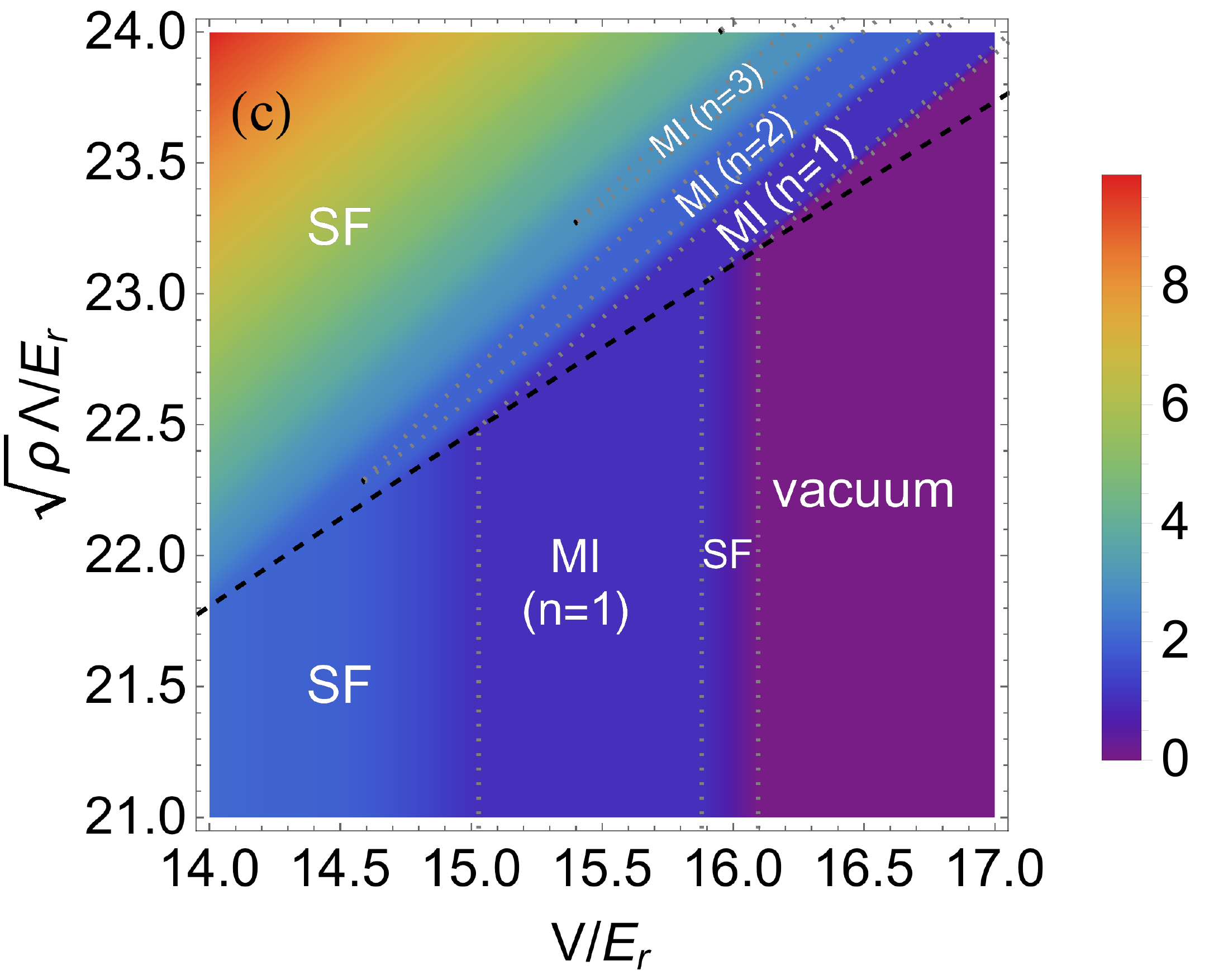}
\caption{
(color online) Phase diagrams between CV and ICV, SF and MI. 
(a,b,c) separately shows the values of \(\alpha_1\)(the real part of \(\alpha\)), \(\beta\) and \(\rho\)
 as functions of of \(V\) and \(\Lambda\) in unit of \(E_\text{r}\)
 (For clear presentation,
  we transform the vertical axis from $\Lambda$ to $\sqrt{\rho}\Lambda$).
The black dashed line depicts the phase boundary between ICV (\(\alpha =0\)) and CV(\(\alpha \neq 0\))~[see Eq.~(\ref{eq:aPB})]. The gray dotted lines depicts the boundaries between SF, MI (\(n=1,2,)\) and vacuum phases~[see Eq.~(\ref{eq:SF-MI})]. 
Here we have taken \(\Omega =100E_\text{r}\), \(\gamma =20E_\text{r}\), \(g=0.1k/m\), and \(\mu=4E_\text{r}\). 
}\label{fig:PD} 
\end{figure}

In the previous section, we have developed the intuitive physical picture and predicted features in the steady state of
an optically trapped quantum gas coupled to a membrane. Below we derive the complete steady-state phase diagram by self-consistently solving Eqs. (\ref{eq:aS}), (\ref{eq:BHMmf}) and (\ref{eq:Smf})
numerically. In particular, we are interested in the steady-states of both the atoms and the membrane, meanwhile the effect of the interplay between the two objects. In determining the steady-state phase diagram of the effective Hamiltonian (\ref{eq:BHMmf}), we stress the existence of two order parameters: $\alpha$ in Eq. (\ref{eq:aEoM}) and $\beta$ in Eq. (\ref{beta}).  In more details,  (i) \(\alpha =0\)  denotes an incoherent vibration (ICV) state of the membrane, whereas \(\alpha \neq 0\) denotes a coherent vibration (CV). (ii)  \(\beta \neq 0\) ( \(\beta =0\)) denotes a superfluid (SF) (Mott-insulator (MI)) state. Depending on the interplay among the two order parameters, we identify four phases in the steady-state phase diagram as follows: (i) $\alpha=0$ and $\beta\neq0$;  (ii) $\alpha\neq0$ and $\beta\neq0$; (iii)  $\alpha=0$ and $\beta=0$ (iv) $\alpha=0$ and $\beta\neq0$.
We point out that Ref. \cite{Mann2018} has only limited into the phase (i) and (ii) and the non-equilibrium from phase (i) to (ii). In this work, we are interested in all above four phases and the corresponding phase boundaries. Our main results are summarized by the diagrams presented in Fig. \ref{fig:PD}, in which we present the values of the order parameters
including \(\alpha _1\)(the real part of \(\alpha\)), \(\beta\) and \(\rho\) as functions of \(V\) and \(\Lambda\), which can be precisely controlled
and tuned in realistic experiments. Other parameters including \(\Omega\), \(\gamma\), \(g\) and \(\mu\) are taken fixed. 

%\subsection*{phase of the membrane}
As stated above, for the steady state of the membrane, its vibration could either be coherent or incoherent, as judged by the order parameter \(\alpha\)
(Due to the interdependency of its real and imaginary parts \(\alpha _1\), \(\alpha _2\), we can focus on \(\alpha _1\)). The dependence of \(\alpha\)
on system parameters is quantified by Eq.~(\ref{eq:aS}), in which the quantity \(S\) has been comprehensively discussed.
% and finally is given by Eq.~(\ref{eq:Ss}).
Thus one can obtain
\begin{equation}
\label{eq:aVL}
%\alpha _1\left\{\frac{4\Lambda ^2 \Omega  }{\tilde{V}\left(\Omega ^2+\gamma ^2\right)}-\exp \left[\left(\frac{E_\text{r}}{\tilde{V}}\right){}^{1/2}\right]\right\}=0,
\alpha _1\left[\frac{4\rho\Lambda ^2 }{\tilde{V}\tilde{\Omega}}-\exp \left(\sqrt{\frac{E_\text{r}}{\tilde{V}}}\right)\right]=0,
\end{equation}
%where \(\tilde{V}\) is given by Eq.~(\ref{eq:Veff}), 
where \(\tilde{V}=\sqrt{V^2+(4\Lambda  \alpha _1)^2}\). Eq.~(\ref{eq:aVL}) implicitly determines the value of \(\alpha _1\). For fixed parameters $\Omega $, $\gamma $, it is solved as a function of \(V\)
and $\Lambda $, as shown in Fig.\ref{fig:PD}(a), where two distinct phase regimes are clearly presented: ICV (\(\alpha =0\)) and
CV(\(\alpha \neq 0\)). 
The CV phase favors for large atom-membrane coupling and weak lattice potential. The phase boundary (see the dashed line in
Fig.\ref{fig:PD}) can also be derived from Eq.~(\ref{eq:aVL}) as 
\begin{equation}
\label{eq:aPB}
\Lambda _c^2=\frac{V\tilde{\Omega}}{4\rho}\exp \left(\sqrt{\frac{E_\text{r}}{V}}\right).
\end{equation}
Notice that the phase boundary for the membrane actually depends on the filling number of the atoms $\rho$.

In addition, in the vicinity of the critical value \(\Lambda _c\) at the CV phase side for fixed $\Omega $, $\gamma $ and \(V\), the critical
behavior of the order parameter is extracted to be
\begin{equation}
\alpha _1\sim \left(\frac{\Lambda -\Lambda _c}{\Lambda _c}\right){}^{1/2}
\end{equation}
with the critical exponent being 1/2. 

%\subsection*{Phase of the bosonic atoms}

The bosonic atoms can be either in the SF phase with the SF order parameter being nonzero, or in the MI phase with the SF order parameter vanishing
and also the filling number being integer. In some simple and clean lattice systems \cite{Greiner2002}, the phase diagram is clearly
determined by the atomic interaction strength and the lattice depth. While in the hybrid system that we are discussing, the physics becomes much
richer due to the coupling between atoms and membrane and the varies tuning parameters. For convenience, we fix again parameters for the membrane
including $\Omega $ and \(\gamma\), and also parameters for the atoms including \(g\) and $\mu $, while investigate the effect of the lattice depth
\(V\) and the atom-membrane coupling. In the regimes for the two different phases of the membrane, the state of the atoms exhibit distinct behaviors,
as shown in Fig.\ref{fig:PD}. 

When the membrane is in the ICV phase, the state of the atoms is irrelevant to the atom-membrane coupling \(\Lambda\), implying that the atoms are
effectively decoupled to the membrane. Due to the vanishing of the order parameter \(\alpha\), their is no other potential than the bare lattice
\(V^{(0)}\), which thus purely determines the phase of the atoms. And its depth \(V\) effectively modulates the hopping \(t\), the Hubbard interaction
\(U\) and also the chemical potential \(\tilde{\mu }\), leading to the SF continuum and MI plateaus. There is also the vacuum phase with the lattice
depth being large sufficiently, meaning trivially no atoms could be supported by the system. The phase boundaries are determined by perturbation theory
\cite{Fisher1989} as
\begin{equation}
\label{eq:SF-MI}
\frac{2 t}{U}=\frac{\left(\rho  -\left.\tilde{\mu }\right/U\right)\left(\left.\tilde{\mu }\right/U+1-\rho \right)}{1+\left.\tilde{\mu }\right/U},
\end{equation}
with $\rho $ being integers for Mott plateaus.

When the membrane is in the CV phase, the state of the atoms is effected not only by the lattice depth \(V\) but also the atom-membrane coupling
\(\Lambda\). Due to the presence of the order parameter \(\alpha\), the extra lattice \(V^{(1)}\) emerges, which constitute into \(V^{\text{eff}}\)
together with the bare lattice \(V^{(0)}\), and modifies the phase diagram of the atoms. The parameters of the effective BHM including \(t\), \(U\),
\(\tilde{\mu }\) are all qualitatively changed, resulting distinct behaviors in the \(V-\Lambda\) plane of the phase diagram. Nevertheless, the phase
boundaries are still determined by Eq.~(\ref{eq:SF-MI}).

\section{Remarks and Conclusions \label{sec:conc}}

Our study is motived by the experimental work \cite{Vochezer2018}, in which they observe the atoms, together with the membrane
will present behavior of collective oscillation, with increasing atom numbers or atom-light coupling, or light-membrane coupling. Theoretically, this inspires
an interesting question as regards the behavior of the quantum gas in the MI regime, although realizing such strong correlated state in a hybrid atom-opto-mechanical system remains experimentally challenging.
While our theory shows that there is an lattice instability, which is contributed by an emergent lattice, whose strength is proportional to the atom number and also
the presence of the coherent membrane mode; as long as some critical values are reached, the lattice minimum will shift, which leads to distinct
behaviors of steady state at zero temperature, as we studied. On the other hand, in the realistic experiment, the couplings are gradually turned
on as time increases, until the extra lattice emerges, leading to a shift in the lattice minimum. Then the atoms will exhibit collective
dynamics, similar to a dipole oscillation.

%\subsection*{Conclusions}
In conclusion, 
we have studied a hybrid atomic and optomechanical system realized experimentally, and found four different
steady-state phases at the mean-field level, characterized by two order parameters: the atomic superfluidity and the vibrational mode of the membrane.
Phase transitions are detuned by the lattice depth and coupling of atom-membrane. 
We analyze that the atom-membrane coupling, which is proportional
to the expectation value of the membrane mode,
 serves as an dynamical lattice for atoms. For small value of the coupling strength, the membrane vibration is incoherent,
  such that the phase of atoms, either being SF or MI, is simply driven by the bare lattice depth. 
As the increase of the coupling strength,
the membrane vibration becomes coherent, and the emergent lattice qualitatively changes the picture of the SF-MI transition. 
The whole phase diagram we obtained will help to gain comprehensive understanding of this hybrid system.
% and will stimulate further investigations.
%Beyond mean-field , quantum fluctuations, like that found in intra-cavity lattice bosons and cavity mode, \cite{Chen2017,Blab2018}.

The experimental realization of the phase diagram studied in this work requires control of two parameters: the lattice strength $V$ and the effective atom-membrane coupling $\lambda$. 
With the state-of-the-art technology \cite{Vochezer2018}, the variation of $V$ and $\lambda$ can be reached by tuning
the laser power and cavity finesse. Moreover,  one can adjust the value of $\lambda$ independent on $V$ by applying a
second laser which is slightly misaligned with the first one generating an optical lattice of the same periodicity but shifted by $\pi/2$. We hope this work can contribute to the ongoing
experiments of quantum gases in a lattice coupled to a membrane.

\section{Acknowledgments}

We thank M. Reza Bakhtiari, and Yu Chen for inspiring discussion and Zhigang Wu for careful reading our manuscript and helpful suggestions.
This work is supported by the National Natural Science Foundation of China (Grant No.11604300) and Key Projects of the Natural Science Foundation of China (Grant No. 11835011).

\bibliography{AOM}
\end{document}